\documentclass[prb,twocolumn,showpacs]{revtex4}
\usepackage{graphicx}
\usepackage{amsmath}
\usepackage{amssymb}

\def\fsl#1{\setbox0=\hbox{$#1$} 
   \dimen0=\wd0 
   \setbox1=\hbox{/} \dimen1=\wd1 
   \ifdim\dimen0>\dimen1 
      \rlap{\hbox to \dimen0{\hfil/\hfil}} 
      #1 
   \else 
      \rlap{\hbox to \dimen1{\hfil$#1$\hfil}} 
      / 
   \fi} %

\newcommand{\sgn}{\mbox{sgn}}

\newcommand{\VEV}[1]{\langle #1 \rangle}

\newcommand{\be}{\begin{equation}}
\newcommand{\ee}{\end{equation}}
\newcommand{\ba}{\begin{eqnarray}}
\newcommand{\ea}{\end{eqnarray}}

\begin{document}

\title{Excitonic gap, phase transition, and quantum Hall effect in graphene: strong-coupling regime}

\date{\today}


\author{V.P. Gusynin}

\affiliation{Bogolyubov Institute for Theoretical Physics, 03143,
Kiev, Ukraine}

\author{V.A. Miransky}

\altaffiliation[On leave from ]{ Bogolyubov Institute for
Theoretical Physics, 03143, Kiev, Ukraine}

\affiliation{ Department of Applied Mathematics, University of
Western Ontario, London, Ontario N6A 5B7, Canada\\
and Yukawa Institute for Theoretical Physics, Kyoto University, Kyoto
606-8502, Japan}

\author{S.G. Sharapov}

\affiliation{Department of Physics and Astronomy, McMaster
University, Hamilton, Ontario L8S 4M1, Canada}

\author{I.A.~Shovkovy}

\altaffiliation[On leave from ]{ Bogolyubov Institute for
Theoretical Physics, 03143, Kiev, Ukraine}

\affiliation{Department of Physics, Western Illinois University,
Macomb, IL 61455, USA}

\date{\today}

\begin{abstract}
We suggest that physics underlying the recently observed removal of
sublattice and spin degeneracies in graphene in a strong magnetic
field describes a phase transition connected with the generation of
excitonic and spin gaps. The strong-coupling regime is described
using a phenomenological model with enhanced Zeeman splitting (spin
gap) and excitonic gaps. The experimental form of the Hall
conductivity $\sigma_{xy}$ with the additional $\nu= 0, \pm1$ plateaus
is reproduced. The form of $\sigma_{xy}$ in the case of a strong-coupling
regime with no enhanced Zeeman splitting is also discussed.
\end{abstract}


\maketitle

\section{Introduction}
\label{sec:intro}

In a recent paper Ref.~\onlinecite{Gusynin2006catalysis}, hereafter
referred to as {\sl I}, we considered how the new plateaus in the
Hall conductivity of graphene discovered by Y.~Zhang {\em et
al.}\cite{Zhang2006PRL} develop, assuming the weak-coupling regime
of the magnetic catalysis scenario. We have become aware of the
experiments\cite{Geim-private} showing that the $\nu=0$  plateau in
the Hall conductivity exists even at rather high temperatures. This
fact indicates the relevance of a strong-coupling regime. It is the
purpose of the present work to complete the theoretical analysis of
{\sl I} by considering also the strong-coupling regime.

A graphite monolayer, or {\em graphene,\/} has become a new exciting topic in
physics of two dimensional electronic systems.\cite{Novoselov2004Science,rev1,rev2}
A qualitatively new feature of graphene is that the eigenfunctions and the
eigenvalues of the low energy quasiparticle excitations are described by the
relativistic $2+1$ dimensional Dirac theory. The spinor structure of the wave
functions is a general consequence of the honeycomb lattice
structure of graphene with two carbon atoms per unit
cell.\cite{Semenoff1984PRL,DiVincenzo1984PRB} When a magnetic
field is applied, noninteracting Dirac quasiparticles occupy Landau levels
(LLs) with the energies
\begin{equation}
\label{E_n}
\begin{split}
& E_n = \sgn (n) \sqrt{2 |n| \hbar v_F^2 |eB|/c} \\
& \approx 424 \, \sgn(n) \sqrt{|n|}  \sqrt{B[\mbox{T}]} \, \mbox{K} ,
\quad n =0, \pm 1,\pm 2, \ldots
\end{split}
\end{equation}
Here to estimate the energies $E_n$, the value of the Fermi velocity
$v_F = 10^{6} \mbox{m/s}$ was used, and the magnetic field $B$
orthogonal to the graphene's plane is given in Tesla. Several
anomalous properties of graphene are attributed to the presence in
the spectrum (\ref{E_n}) of the $n=0$ field independent lowest
Landau level (LLL). In the diagonal conductivity, the anomaly
manifests itself as the phase shift of $\pi$ of the quantum magnetic
oscillations expected theoretically both using the semiclassical
quantization condition for the quasiparticles with a linear
dispersion\cite{Mikitik1999PRL} and from a microscopical
calculation valid both for the massless and massive Dirac
fermions.\cite{Sharapov2004PRB,Gusynin2005PRB} Accordingly, in the Hall
conductivity the anomaly results in the anomalous integer quantum Hall
(QH) effect with the plateaus at the filling factors $\nu = \pm
4(|n|+1/2)$.\cite{Zheng2002PRB,Gusynin2005PRL,Gusynin2006PRB,Peres2006PRB}
These theoretical conceptions allowed to identify unambiguously the
Dirac quasiparticles in the two independent
experiments.\cite{Geim2005Nature,Kim2005Nature} At present a lot of
studies on graphene are concentrated on phenomena whose understanding
demands going beyond unconventional but yet rather simple physics of
noninteracting Dirac quasiparticles.

On the experimental side, for a magnetic field $B \gtrsim 20\, \mbox{T}$,
the appearance of additional QH plateaus with the filling factors $\nu =0,
\pm1, \pm 4$ was reported in Ref.~\onlinecite{Zhang2006PRL}. The
theoretical studies of these additional plateaus in graphene can be
divided into four classes.

\noindent
(i) Fractional QH effect. Although there is no experimental evidence
for such an effect in graphene so far, there are a few theoretical papers,
where this possibility is
discussed.\cite{Peres2006PRB,Apalkov2006,Toke2006,Khveshchenko2006FQHE}

The remaining three scenario consider various possibilities of breaking
the $U(4)$ symmetry of the non-interacting Hamiltonian of graphene.
While this symmetry remains intact in the presence of a long-range
Coulomb interaction and at a nonzero chemical potential $\mu$, it is
explicitly broken in the presence of the Zeeman and some types of
short-range interactions.

\noindent
(ii) A model with local (on-site) interactions which explicitly
break the $U(4)$ symmetry was considered
in Ref.~\onlinecite{Alicea2006PRB}.

\noindent
(iii) An analogy between the four-fold degeneracy of LLs in graphene
associated with the $U(4)$ symmetry and the $SU(4)$ ferromagnetism
studied previously in the bilayer quantum Hall systems\cite{foot1} is
exploited in Refs.~\onlinecite{Nomura2006PRL,Goerbig2006}. In this
QH ferromagnetism scenario the QH plateaus with {\it all} integer values
of the filling factor $\nu$ occur. The current experimental
data,\cite{Zhang2006PRL} however, do not seem to support the
existence of the plateaus with $\nu=\pm3,\pm5,\ldots$.

\noindent
(iv) The magnetic catalysis scenario was considered in {\sl I}
and in Refs. \onlinecite{Herbut2006,Ezawa2006} .
It is based on the phenomenon of the
electron-hole (fermion-antifermion) pairing
in a magnetic field revealed in field theory\cite{Gusynin1995PRD}
and the analysis in Refs.~\onlinecite{Khveshchenko2001PRL,
Gorbar2002PRB,Gorbar2003PLA,Khveshchenko2004nb},
where this phenomenon was considered in graphene. (This
analysis was originally inspired by the early experiments in highly
oriented pyrolytic graphite.\cite{kopel1}) In Refs.~\onlinecite{Khveshchenko2001PRL,
Gorbar2002PRB,Gorbar2003PLA,Khveshchenko2004nb},
the spontaneous breakdown of the $U(4)$ symmetry by generating a
dynamical excitonic gap $\Delta$ was considered. A new development
for this scenario suggested in {\sl I} was to fit the data of
Ref.~\onlinecite{Zhang2006PRL} by including both the Zeeman term
and the excitonic gap. The central feature of this scenario is
that the only plateaus in the Hall conductivity $\sigma_{xy}$ are those with
$\nu = 0, \pm 1$ and $\nu = \pm 2k$ ($k=1, 2,\ldots $), i.e., the plateaus
observed in experiment.\cite{Zhang2006PRL}. While in {\sl I}
the excitonic gap is produced by the Coulomb interactions, local
(on-site) interactions are used for this purpose in Ref.
\onlinecite{Herbut2006}. A dynamics relating to the magnetic catalysis
scenario was considered in Ref. \onlinecite{Fuchs2006}, in which an
excitoniclike gap is produced by electron-phonon interactions.

As was emphasized in {\sl I}, while the plateau $\nu =0$ could even appear
either due to the spin splitting or the excitonic gap
$\Delta$ alone, the plateaus $\nu = \pm 1$ arise
only if both the spin splitting and the gap $\Delta$ are non-vanishing.
In other words, the  plateaus $\nu = \pm 1$ are generated by the
dynamics which completely removes the $U(4)$ degeneracy of the LLL.
This can for example be seen explicitly from the expression for the Hall
conductivity due to the $n=0$ Landau level in the clean limit (i.e., in the
limit of the vanishing scattering rate of quasiparticles):
\begin{equation}
\label{sigma_xy-ideal}
\begin{split}
& \sigma_{xy}=-\frac{e^2}{h}{\rm sgn}(eB)\\
& \times \left[{\rm sgn}(\mu_{+}) \theta(|\mu_{+}|-\Delta)+{\rm
sgn}(\mu_{-})\theta(|\mu_{-}|-\Delta)\right],
\end{split}
\end{equation}
where $\mu_{\pm} = \mu \pm E_Z$ with $E_Z$ being the Zeeman energy.
The fitting procedure in {\sl I} is heavily based on the assumption that the
Zeeman energy is
\begin{equation}
\label{Zeeman-usual} E_Z = \frac{g_L}{2} \mu_B B \simeq
0.67\, B[\mbox{T}] \, \mbox{K} ,
\end{equation}
where $\mu_B = e\hbar/(2 m c)$ is the Bohr magneton and the Lande
factor in graphene is $g_L \simeq 2$. For typical strengths of the
magnetic field used in the experiment, $B\lesssim 45 \,\mbox{T}$, the
Zeeman energy (\ref{Zeeman-usual}) is $E_Z\lesssim 30 \,\mbox{K}$.
By combining this observation with the fact that the $\nu=0$ and
$\nu = \pm 1$ plateaus have comparable widths,\cite{Zhang2006PRL}
it was concluded that the excitonic gap $\Delta$ is of the same order
as $E_Z$. Indeed, as seen from Eq.~(\ref{sigma_xy-ideal}), it is the
interplay between these two energy scales, $\Delta$ and $E_Z$, that
determines the size of the lowest plateaus. The required magnitude
(of several dozens of Kelvin) for the excitonic gap corresponds to the
weak-coupling regime. In {\sl I}, the best fit was found at $g =
e^2/(\epsilon \hbar v_F)\simeq 0.07$, where $\epsilon$ is the dielectric
constant of the medium. An immediate implication of such a weakly
coupled dynamics is that the typical magnitude of the critical temperature,
at which the gap disappears, is of the same order of a dozen of Kelvin.

In part, the motivation for the present work is the new experimental
data\cite{Geim-private} suggesting that the $\nu=0$ plateau,
observed for strong magnetic fields, may persist even at rather high temperatures.
This alone would suggest the relevance of the strong-coupling regime with
$g\gtrsim 1$. Moreover, as will be shown in Sec.~\ref{sec:exp} in detail, this
regime is also consistent with the structure of the higher plateaus in the Hall
conductivity reported in Ref. \onlinecite{Zhang2006PRL}.

In contrast to the weak-coupling regime, the dynamics at strong
coupling is non-perturbative and, therefore, permits no rigorous
quantitative analysis. In addition, the simple type of the dynamics
discussed in {\sl I} is not sufficient to fit the experimental data
in Ref. \onlinecite{Zhang2006PRL}.
This is because of a very large hierarchy between the energy scales
set by the Zeeman splitting (\ref{Zeeman-usual}) and the excitonic
gap at strong coupling. In a consistent approach, as will be
discussed below, an anomalous enhancement of the Zeeman-like
splitting will be required.

Because of the difficulties due to the non-perturbative dynamics at
strong coupling, we use a phenomenological approach in this paper.
Therefore, most conclusions of the analysis below should be viewed
only as qualitative. As we shall see, however, many of them are
likely to be very robust. For example, the critical temperature for
the excitonic condensate as well as for the related appearance of
the plateaus $\nu = 0, \pm 1$ should be of the same order as the
excitonic gap, i.e., a few hundred Kelvin.

The paper is organized as follows. In Sec.~\ref{sec:exp}, we analyze
the experimental data of Ref.~\onlinecite{Zhang2006PRL} in detail,
and extract the constraints on the dynamics to be implemented in a
phenomenological model. The model itself is introduced in
Sec.~\ref{sec:model}. In Sec.~\ref{sec:results}, we present a fit of
the experimental data for the $\nu = 0,\pm 1$ as well as the higher
plateaus, and underline the specific features of the strong-coupling
regime. In Sec. \ref{sec:noZeeman}, for completeness and better
understanding the role of the Zeeman splitting,
we consider the dynamics in the strong-coupling regime with no
enhanced Zeeman splitting. In this case, besides the
standard sequence $\nu = (4n+2)$,
the only additional plateau is that with $\nu=0$.
In Sec.~\ref{sec:concl}, the main results of the paper are
summarized.

\section{Analysis of the experimental data}
\label{sec:exp}

The low-energy quasiparticles excitations in graphene are described
in terms of a four-component Dirac spinor $\Psi_\sigma^T= \left(
\psi_{KA\sigma},\psi_{KB\sigma},\psi_{K^\prime B\sigma},
\psi_{K^\prime A\sigma}\right)$. This spinor combines the Bloch
states with spin $\sigma=\pm1$ on the two different sublattices
($A,B$) of the hexagonal graphene lattice and with momenta near
the two inequivalent points ($K,K^\prime$) at the opposite corners
of the two-dimensional Brillouin zone.

The free, low-energy quasiparticle Hamiltonian for zero carrier density
(or $\mu=0$) and in absence of the Zeeman splitting can be recast in
a relativistic form,
\be
H_0=-iv_F\int d^2{\bf
r}\,\overline{\Psi}_\sigma\left(\gamma^1\hbar\nabla_x+
\gamma^2\hbar\nabla_y\right)\Psi_\sigma, \label{free-hamiltonian}
\ee
where $\overline{\Psi}_\sigma=\Psi^\dagger_\sigma\gamma^0$ is the
Dirac conjugated spinor and summation over spin $\sigma$ is understood.
Notice that the Fermi velocity $v_F\approx 10^6\,\mbox{m/s}$ plays the
role of the speed of light. In Eq.~(\ref{free-hamiltonian}),
$\gamma^\nu$, $\nu=0,1,2$, are $4\times4$ gamma matrices
belonging to a reducible representation of the Dirac algebra:
$\gamma^\nu=\tilde{\tau}_3\otimes (\tau_3,i\tau_2,-i\tau_1)$,
where the Pauli matrices $\tilde{\tau},\tau$ act in the subspaces of
the valley ($K,K^\prime$) and sublattices ($A,B$) indices,
respectively. The matrices satisfy the usual anticommutation
relations $\left\{\gamma^\mu,\gamma^\nu\right\}=2g^{\mu\nu}$,
$g^{\mu\nu}=(1,-1,-1)\,, \mu,\nu=0,1,2$. The orbital effect of
a magnetic field $\mathbf{B}$ applied perpendicular to the
graphene plane is included via the covariant derivative
${\pmb \nabla}={\pmb \partial}+(ie/\hbar c){\bf A}$.

The explicit form of the interaction Hamiltonian and the effective
action are derived and discussed in {\sl I}. We describe their
features in the strong-coupling regime below.

\subsection{Relationship between the gate voltage and $\mu$ }

In experiments\cite{Novoselov2004Science,Geim2005Nature,Kim2005Nature,Zhang2006PRL}
the density of carriers (or carrier imbalance $\rho$ which is the
difference between the densities of electron and holes) is tunable
by the gate voltage $V_g$ applied to the Si substrate of a graphene
device. The measurements of the Hall coefficient and a (consistent with
them) theoretical estimate give the relationship\cite{Geim2005Nature}
\begin{equation}
\rho = \alpha (V_g-V_0), \qquad \alpha \approx 7.3 \times 10^{10}\,
\mbox{cm}^{-2}\, \mbox{V}^{-1},
\label{shift}
\end{equation}
where the shift $V_0$ is attributed to the shift of the Dirac point
due to an unintentional doping.\cite{Novoselov2004Science} In our
analysis of the experimental data,\cite{Zhang2006PRL} the value of
the shift is in the range from $0.8\, \mbox{V}$ to $5.8\, \mbox{V}$.

On the theory side, the variation of the density of carriers in the system
is modeled through adding to the Hamiltonian (\ref{free-hamiltonian})
the term $-\mu \overline{\Psi}_\sigma \gamma^0\Psi_\sigma =
-\mu \Psi^{\dagger}_\sigma\Psi_\sigma$ with a tunable chemical
potential $\mu$. In order to make a connection between the
experiment and the theory, one needs to know the relationship
between the chemical potential $\mu$ and the carrier imbalance
$\rho$.

We have examined the data of Ref.~\onlinecite{Zhang2006PRL} relying
on the following two natural {\em assumptions}:

\begin{enumerate}
\item[(i)] the middle point of a step between two neighboring plateaus
is associated with the value of the chemical potential in the middle of
a smeared Landau level, and

\noindent
\item[(ii)] the middle of a plateau is associated with the value of the
chemical potential in the middle of the energy gap between two
Landau (sub)-levels.
\end{enumerate}

The meaning of these two assumptions might be easier to grasp by
referring to the correspondence between the quasiparticle energy
spectrum and the structure of the Hall conductivity plateaus as shown,
e.g., in Fig.~\ref{spectrum} below.

\begin{figure}
\begin{center}
\includegraphics[width=.48\textwidth]{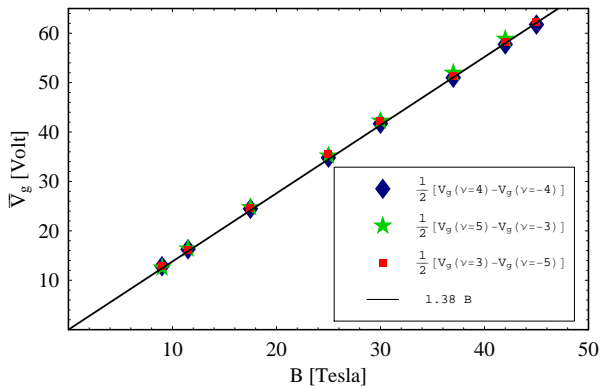}
\caption{(Color online) The compilation of the experimental data from
Ref.~\onlinecite{Zhang2006PRL} for the gate voltage $V_g(\nu)$
differences, $1/2[V_g(\nu=4) - V_g(\nu=-4)]$, $1/2[V_g(\nu=5) -
V_g(\nu=-3)]$ and $1/2[V_g(\nu=3) - V_g(\nu=-5)]$ as a function of
the magnetic field $B$. } \label{iterpolation1}
\end{center}
\end{figure}

\begin{figure}
\begin{center}
\includegraphics[width=.48\textwidth]{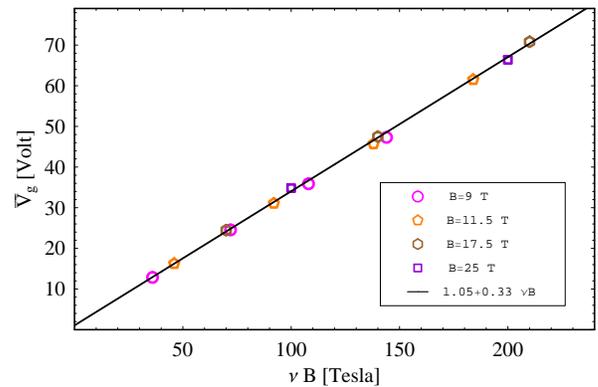}
\caption{(Color online) The compilation of the experimental data from
Ref.~\onlinecite{Zhang2006PRL} for the voltage differences
$1/2[V_g(\sqrt{|\nu/2| \hbar v_F^2 |eB|/c}) - V_g(-\sqrt{|\nu/2|
\hbar v_F^2 |eB|/c})]$ that correspond to the first few observed
sets of the filling factors as a function of $\nu B$ for four values
of the field $B=9, 11.5, 17.5, 25 \mbox{T}$.} \label{iterpolation2}
\end{center}
\end{figure}

We note that the structure of the higher plateaus can be used to
extract the dependence $\mu(V_g)$. In the corresponding analysis
it is useful to assume that only the plateaus $\nu =0, \pm 1, \pm 2$
(connected with the LLL) are affected by the excitonic gap. This
qualitative feature is a rigorous outcome in the weak-coupling
regime of {\sl I}. If it holds also at strong coupling, which is at
least plausible,\cite{Gusynin1995PRD} the dynamics of the
exciton condensation would not contaminate the structure of
the higher plateaus in the Hall conductivity. Then, for the filling
factors $\nu=\pm3,\pm4,\pm5$, one derives the following
relations:
\begin{subequations}
\label{assumptions}
\begin{align}
& V_{g}(\nu = \pm 4) = V_g(\mu = E_{\pm 1}), \label{ii-a} \\
& V_{g}(\nu = \pm 3) = V_g(\mu = E_{\pm 1} \mp E_Z), \label{i-b}\\
& V_{g}(\nu = \pm 5) = V_g(\mu = E_{\pm 1} \pm E_Z) \label{i-c},
\end{align}
\end{subequations}
where Eqs.~(\ref{i-b}), (\ref{i-c}), and Eq.~(\ref{ii-a}) are based
on the assumptions (i) and (ii), respectively [see also
Fig.~\ref{spectrum} below]. In the analysis it is reasonable to
accept that the Zeeman-like energy $E_Z$, characterizing the Landau
level splitting, is considerably smaller than $E_1$ (i.e., $E_Z \ll
E_1$) even if $E_Z$ is substantially enhanced compared to the
ordinary Zeeman energy in Eq.~(\ref{Zeeman-usual}). As we will see
in Subsec.~IIB and IIC below, this hierarchy of scales is indeed
consistent with the strong-coupling regime: while $E_Z \simeq 100\,
\mbox{K}$, the energy $E_1 \simeq 1000\, \mbox{K}$. The results of
the analysis are summarized in Figs.~\ref{iterpolation1} and
\ref{iterpolation2}.

In Fig.~\ref{iterpolation1}, the experimental data for three specific
combinations of the gate voltage differences are plotted as functions
of the external magnetic field. After taking into account the
correspondence between the filling factors and the values of the chemical
potential in Eq.~(\ref{assumptions}), we see that all three gate
voltage differences,
\begin{subequations}
\ba
&&\frac{1}{2}\left[V_g(\nu=4) - V_g(\nu=-4)\right],\\
&&\frac{1}{2}\left[V_g(\nu=5) - V_g(\nu=-3)\right],\\
&&\frac{1}{2}\left[V_g(\nu=3) - V_g(\nu=-5)\right],
\ea
\end{subequations}
determine the average value of the voltage (measured from $V_0$) that,
up to higher order corrections, corresponds to $\mu=E_{1}$, i.e.,
$\mu=424 \sqrt{B\,[\mbox{T}]} \,\mbox{K}$, see Eq.~(\ref{E_n}). (Note
that to linear order the corrections due to Zeeman splitting cancel
in all three combinations.) From Fig.~\ref{iterpolation1}, we see that
the dependence of the voltage on the magnetic field is described almost
perfectly by the following linear fit:
$\overline{V}_g(B) \approx 1.38 \,B[\mbox{T}] \, \mbox{V}$ where
$\overline{V}_g \equiv |V_g-V_0|$ and $V_0$ describes the shift of the
Dirac point, see
Eq.~(\ref{shift}). Therefore, by trading the magnetic field $B$ for the
corresponding chemical potential, i.e.,
$\mu=424 \sqrt{B\,[\mbox{T}]} \,\mbox{K}$,
we get
\be
\overline{V}_g \approx \left(\frac{\mu [\mbox{K}]}{361}\right)^2\, \mbox{V}.
\label{vgvsmu}
\ee
By inverting this relation, we arrive at the following one:
\begin{equation}
\label{V2mu} \mu = \kappa\, \sgn(V_g-V_0) \sqrt{ |V_g-V_0|}, \quad
\kappa \approx 361 \, \mbox{K} \, \mbox{V}^{-1/2}.
\end{equation}
The relations (\ref{vgvsmu}) and (\ref{V2mu}) will be used below to
make a contact between the theory and the experiment. [The relation
(\ref{V2mu}) is shown in Fig.~\ref{spectrum} below by a dotted line.]

Another independent derivation of the relationship between $V_g$ and $\mu$
can be obtained by fitting a sequence of several Landau levels $E_{n}$
at a fixed value of the magnetic field. In Ref.~\onlinecite{Zhang2006PRL},
several lowest plateaus that correspond to the filling factors $\nu = 4n$
(with an integer $n$) were observed for the four lowest values of the field.
In accordance with our assumption (ii), the corresponding values of the
voltage are associated with the chemical potentials in the middle of the
energy gap between the two sublevels of the Zeeman-split $n$th Landau level,
i.e., $\mu=E_{n}$ to linear order.
By making use of Eq.~(\ref{E_n}), the
corresponding chemical potentials are given by $\mu=212\,\sgn(\nu)
\sqrt{|\nu B|\,[\mbox{T}]}\,\mbox{K}$ with $\nu=4n$.

The experimental data for the voltage differences that correspond to the
first few observed sets of the filling factors are plotted versus $\nu B$
in Fig.~\ref{iterpolation2}. This dependence is well approximated by a
linear function, $\overline{V}_g \approx  \left(1.05+0.33\, \nu B\,
[\mbox{T}]\right)\,\mbox{V}$. Note that because of limited data for
low $B$, the value
$1.05\, \mbox{V}$ of a
small intercept term $\alt V_0$ is not reliable and will be
omitted. Then,
by expressing $\nu B$ in terms of the chemical potential (as discussed
in the previous paragraph), we arrive at the relation
$ \overline{V}_g \approx (\mu [\mbox{K}]/369)^2\,\mbox{V}$,
which agrees reasonably well with Eq.~(\ref{vgvsmu}).

Interestingly, the relation
(\ref{V2mu}) with practically the same coefficient $\kappa$ also
follows from the equation $\mu^2 = \pi \hbar^2 v_F^2 |\rho|$ for the
ideal gas of Dirac quasiparticles at $T=B=0$ used in
Refs.~\onlinecite{Gorbar2002PRB, Gusynin2006PRB}. Note however that
the simultaneous use of Eq.~(\ref{V2mu}) and the usual (i.e.,
without enhancement) Zeeman splitting (\ref{Zeeman-usual}) would
lead to inconsistencies in fitting the experimental data. This was
the reason for considering another relation between $\mu$ and $V_g$
at weak-coupling in {\sl I}.

\subsection{Zeeman term}
\label{sec:zeeman}

Now we are in a position to discuss the second, Zeeman term
$\mu_BB\overline{\Psi}\gamma^0\sigma_3\Psi =\mu_BB
\Psi^{\dagger} \sigma_3 \Psi$ which has to be added to the free
Hamiltonian (\ref{free-hamiltonian}) (here the $\sigma_3$ matrix
acts on spin indices). Considering the weak-coupling regime in
{\sl I}, we assumed that the Zeeman splitting is usual as given
by Eq.~(\ref{Zeeman-usual}). There are, however, theoretical
arguments\cite{Nomura2006PRL,Goerbig2006,Abanin2006PRL} that
the Coulomb interaction in the exchange channel may strongly enhance
Zeeman splitting and lead to a spin gap $\Delta_Z$ (expressed through
the condensate $\VEV{ \overline{\Psi}\gamma^0\sigma_3\Psi }$) as
large as a few hundred Kelvin.

\begin{figure}
\begin{center}
\includegraphics[width=.48\textwidth]{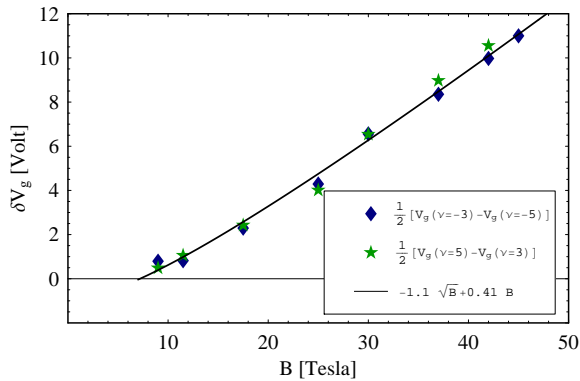}
\caption{(Color online) The compilation of the experimental data from
Ref.~\onlinecite{Zhang2006PRL} for the size of the $\nu=4$ plateau
that determines the Zeeman splitting of the first Landau level.}
\label{1stLLzeeman}
\end{center}
\end{figure}

Using the experimental data of Ref.~\onlinecite{Zhang2006PRL}, the
energy of the Zeeman splitting can be found from the size of
$\nu = \pm 4$ plateau whose appearance is attributed to the lifting
of the spin degeneracy of the $n=\pm1$ Landau level. We estimate
the size of this plateau by extracting the following gate voltage
differences:
$\delta V_g \equiv 1/2[V_g(\mu =E_1 + E_Z)-V_g(\mu =E_1 - E_Z) ]
\approx 1/2[V_g(\mu =E_{-1} + E_Z) -V_g(\mu=E_{-1} - E_Z) ]$.
The results are presented in Fig.~\ref{1stLLzeeman}. We find that
the size of $\sigma_{xy} = \pm 4e^2/h$ plateau is reasonably well
fitted by the dependence
\be
\delta V_g(B) \approx (- 1.1 \sqrt{B [\mbox{T}]}
+0.41 B[\mbox{T}])\,\mbox{V}
\label{DeltaVg}
\ee
for $9\, \mbox{T} < B <45\, \mbox{T}$. On the other hand, if
$\delta \mu$, corresponding to $\delta V_g$, is reasonably small,
one can differentiate expression (\ref{vgvsmu}) to obtain the
relation $\delta V_g \approx 2 \mu [\mbox{K}]\, \delta \mu
[\mbox{K}]/(361)^2\,\mbox{V}$
(here the differentials $dV_g$ and $d\mu$ were replaced by
$\delta V_g$ and $\delta \mu$, respectively). Now,
substituting the chemical potential
$\mu = E_{1} = 424 \sqrt{B [\mbox{T}]}\, \mbox{K}$ and
the splitting $\delta \mu =E_Z$ in this expression
for  $\delta V_g$
and comparing it with Eq.~(\ref{DeltaVg}),
we arrive at the relation
\be \label{Zeeman-enh} E_Z = (63 \sqrt{B [\mbox{T}]} - 169)\,\mbox{K}.
\ee
This is 3 to 8 times larger than the usual Zeeman energy
(\ref{Zeeman-usual}) for $9\, \mbox{T} < B <45\, \mbox{T}$, providing
another argument in favor of strong coupling.
The large difference between the enhanced Zeeman
energy(\ref{Zeeman-enh}) and conventional Zeeman energy
(\ref{Zeeman-usual}) is likely caused by dynamical
effects.\cite{Nomura2006PRL,Goerbig2006,Abanin2006PRL} In general,
these effects should be sensitive to the temperature and chemical
potential, so that $E_Z$ is also a function of $T$ and $\mu$. Here,
however, we restrict ourselves by the dependence $E_Z$ on $B$.

\subsection{Gap terms and the $\nu = \pm 1$ plateaus}
\label{sec:gap-exp}

As is clear from Eq.~(\ref{sigma_xy-ideal}), the $\nu = \pm 1$ plateaus
of the Hall conductivity observed in Ref.~\onlinecite{Zhang2006PRL}
can exist only if there is Zeeman splitting {\it and} the spectrum
(\ref{E_n}) is gapped, viz.
\begin{equation} \label{E_n-gap}
\begin{split}
& E_n = \sgn(n) \sqrt{2 |n| \hbar v_F^2 |eB|/c + \Delta^2}.
\end{split}
\end{equation}
In this case the four-fold degeneracy of the LLL ($n=0$) is completely
lifted.

Considering the dependence of $\sigma_{xy}$ on $\mu$ in
Eq.~(\ref{sigma_xy-ideal}), one can see that while the size of
the $\nu = \pm 1$ plateaus is $\delta\mu_{1}\simeq 2 E_Z$,
the size of the 0th plateau is $\delta\mu_{0}\simeq 2(\Delta-E_Z)$.
The experiment in Ref.~\onlinecite{Zhang2006PRL} measures
the dependence $\sigma_{xy}(V_g)$ and indicates that the sizes of the
0th and $\nu = \pm 1$ plateaus (as a function of $V_g$) are comparable.
Using Eq.~(\ref{V2mu}), we estimate the $V_{g}$-size of the $\nu = \pm 1$
plateaus as $4\Delta E_Z/\kappa^2$ (in Volts), while the size of
the 0th plateau is $2(\Delta-E_Z)^2/\kappa^2$ (in Volts). By
simply requiring these two quantities to be equal, we derive
$\Delta \simeq (2+\sqrt{3})E_Z$. By making use of the
Zeeman energy (\ref{Zeeman-enh}) extracted from experiment,
we obtain $E_Z \approx 176 \,\mbox{K}$ at a moderately strong
field $B = 30 \, \mbox{T}$. Therefore, a typical value of
the excitonic gap is $\Delta \simeq 657 \, \mbox{K}$.

It was assumed in {\sl I} that the gap $\Delta$  in
Eq.~(\ref{E_n-gap}) is driven by the singlet excitonic order
parameter
\ba \label{singlet-OP} \VEV{\overline{\Psi}\sigma_0\Psi}
&=&\sum_{\sigma = \pm 1}\VEV{ [\psi_{KA\sigma}^\dagger
\psi_{KA\sigma} + \psi_{K^\prime A\sigma}^\dagger \psi_{K^\prime
A\sigma}
\nonumber\\
&&- \psi_{KB\sigma}^\dagger \psi_{KB\sigma} - \psi_{K^\prime
B\sigma}^\dagger \psi_{K^\prime B\sigma}]}, \ea
where we explicitly wrote unit matrix $\sigma_0$ acting on spin indices to
underline the singlet character of this order parameter.
The Zeeman interaction, however, may favor\cite{Khveshchenko2004nb,Herbut2006}
the triplet order parameter, $\VEV{\overline{\Psi}\pmb{\sigma}\Psi}$, where
the vector $\pmb{\sigma} = (\sigma_1, \sigma_2, \sigma_3)$ made from Pauli
matrices acting on spin indices. Therefore, in addition to the singlet
order parameter (\ref{singlet-OP}), one may also include the triplet
order parameter\cite{foot2}
\ba \label{triplet-OP} \VEV{\overline{\Psi}\sigma_3 \Psi}&=&
\sum_{\sigma = \pm 1}\sigma \VEV{ [\psi_{KA\sigma}^\dagger
\psi_{KA\sigma} + \psi_{K^\prime A\sigma}^\dagger \psi_{K^\prime
A\sigma}
\nonumber\\
&&- \psi_{KB\sigma}^\dagger \psi_{KB\sigma} - \psi_{K^\prime
B\sigma}^\dagger \psi_{K^\prime B\sigma}]}.
\ea
In the language of symmetry, the Zeeman term explicitly breaks the $U(4)$
symmetry down to the $U(2)_c \times U(2)_d$ (see Appendix~C of {\sl I}).
[Note that the condensates $\VEV{\overline{\Psi}\sigma_1 \Psi}$ and
$\VEV{\overline{\Psi}\sigma_2 \Psi}$ would
spontaneously break the $SO(2)$ symmetry
of the in-plane rotations and will not be considered here.] The
dynamical generation of the gaps connected with order parameters
(\ref{singlet-OP}) and (\ref{triplet-OP})
leads to the spontaneous breakdown of the $U(2)_c \times U(2)_d$ symmetry
down to the abelian $U(1)_1 \times U(1)_2 \times U(1)_3 \times U(1)_4$
one.
There might exist also other gaps that break the $U(2)_c \times U(2)_d$
symmetry. Therefore, the problem is very complicated in general due to the
potential possibility of many competing order parameters.

\section{Phenomenological model}
\label{sec:model}

There are several ways to tackle the problem of competing order parameters
in graphene. The best way would be to include all relevant interactions and
study the instability of the system with respect to the formation of all
possible condensates. An approach of this type was recently attempted in
Ref.~\onlinecite{Herbut2006} in a weak-coupling regime. The applicability
of such an approach is extremely limited at strong coupling, however.

The second option is to use a phenomenological expression for the
Zeeman splitting as an input for the thermodynamic potential and
minimize the latter only with respect to the gaps of interest.
In {\sl I}
we used the latter approach, taking as an input the usual Zeeman
splitting (\ref{Zeeman-usual}) and considering the thermodynamic
potential with the singlet excitonic gap $\Delta$ in the weak-coupling
regime with $g \simeq 0.07$.

The analysis is reliable in this regime, but the results do not describe
all the features of the Hall conductivity in graphene quantitatively. In fact,
the analysis of the data presented in Sec.~\ref{sec:exp} led us to the conclusion
that the weak-coupling regime is most likely improbable. Therefore, here we
insist on the strong-coupling regime to describe the physics of graphene. Our
argument is the following. The singlet gap $\Delta$ in {\sl I} was related to the
Landau scale $L(B) = \sqrt{\hbar v_F^2|eB|/c}$ via the solution of the gap
equation, $\Delta = b L(B)$, where the dimensionless parameter $b$ in the
LLL approximation is given by
\be \label{b} \quad
b=\frac{g}{\sqrt{2}}\int\limits_0^\infty\frac{dk\,
e^{-k^2}}{1+k\chi_0}, \ee
with $\chi_0\simeq 0.56\sqrt{2}\pi g$. Since the estimate for the value
of $\Delta$ obtained in Sec.~\ref{sec:exp} is by an order of magnitude
larger than in {\sl I}, the corresponding parameter $b=0.4$ is also 10
times larger than in {\sl I}. Then, using Eq.~(\ref{b}), we estimate
that the value of the coupling constant is $g \sim 1.56$ which implies
a strong-coupling regime.

In this regime, unfortunately, there exist no reliable schemes for
treating the pairing dynamics quantitatively. As was emphasized in
Ref.~\onlinecite{Gorbar2002PRB}, the problem is due to non-decoupling
of the low-energy dynamics on the LLL from the dynamics on higher Landau
levels.

A general insight into the dynamics of magnetic catalysis at
strong-coupling can be gained from the original work\cite{Gusynin1995PRD}
in models with short range (contact) interactions in the leading order
of $1/N$ expansion, where $N$ is the number of fermion ``flavors" ($N$
equals 2 in graphene). The dynamical picture following from that analysis
is the following. While an external magnetic field strongly enhances the
pairing dynamics, a nonzero density of carriers tends to suppress pairing.
Moreover, for a wide range of parameters, the gap closes at the critical
density that corresponds to filling the LLL. In terms of the chemical
potential, this gives the critical value, $|\mu_c|=\Delta$ in the absence
of the Zeeman splitting,\cite{Gorbar2002PRB} and $|\mu_c|=\Delta+E_Z$ when
the Zeeman splitting is relevant.

In addition to the exciton condensation in graphene, one should also
account for the anomalously large Zeeman splitting (\ref{Zeeman-enh}),
which is likely to have a similar dynamical origin. As is discussed in {\sl I}
and in Sec.~\ref{sec:exp} above, it is the only way to remove completely
the four-fold degeneracy of the lowest Landau level necessary for
the explanation of the experimentally observed $\nu=1$ plateau in the
Hall conductivity.

Here, therefore, we implement all the details of the fermion pairing
dynamics by making use of a simple phenomenological approach. In
particular, we will use the simplest possible ansatz for the dynamically
generated gaps $\Delta_{\pm}$ for the spin up and down states, which
capture the essential features of the magnetic catalysis phenomenon
discussed above. We write
\begin{equation}
\begin{split}
\label{gap-ansatz}
\Delta_{+}(B,\mu)   &=  \Delta(B,\mu_+ ) ,\\
\Delta_{-} (B,\mu)   &=  \Delta(B,\mu_- ) ,
\end{split}
\end{equation}
where, as already defined in the text after Eq.~(\ref{sigma_xy-ideal}),
$\mu_{\pm} = \mu \pm E_Z(B)$, with the Zeeman energy $E_{Z}(B)$ given
in Eq.~(\ref{Zeeman-enh}). Note that the triplet channel is taken into
account by considering the two different gaps $\Delta_{\pm}$ corresponding
to the order parameters $1/2 \VEV{\overline{\Psi}(\sigma_0 \pm \sigma_3)\Psi}$,
respectively. The magnetic catalysis dynamics is mainly realized at the
two Fermi surfaces, $\mu_{\pm}=0$, connected with the spin up and down
quasiparticles, and leads to the gaps of approximately the following form:
\begin{widetext}
\begin{equation}
\label{gap-ansatz-main}
\Delta(B,\mu_{\pm}) = \frac{\Delta_0}{\pi} \theta
(B-B_c)\sqrt{\frac{B}{B_c}-1}
\left[\arctan\left(\frac{\mu_{\pm}+\mu_c(B)}
{\gamma}\right)-\arctan\left(\frac{\mu_{\pm}-\mu_c(B)}{\gamma}\right)
\right].
\end{equation}
\end{widetext}
Here $B_c$ is the critical field, $\gamma$ is the LLL width (or
quasiparticle scattering rate), and the dependence of the critical
chemical potential on the field $B$ is
\begin{equation}
\label{mu-ansatz}
 \mu_{c}(B) = \Delta_0 \theta
(B-B_c)\sqrt{\frac{B}{B_c}-1}.
\end{equation}
The expression (\ref{gap-ansatz-main}) incorporates the following key
features of the magnetic catalysis dynamics:

\begin{itemize}
\item[(i)] In view of expression (\ref{mu-ansatz}), the gap is negligible
for chemical potentials larger than the critical value $\mu_c(B)$ which
corresponds to the filling of the LLL, i.e., $\mu_c \simeq \Delta(B,\mu)|_{\mu=0}$
(see Ref.~\onlinecite{Gusynin1995PRD,Gorbar2002PRB}).

\item[(ii)] In accordance with previous studies,\cite{Gorbar2002PRB}
the magnetic catalysis occurs only when the field exceeds a critical
value $B_c$.

\item[(iii)] The effects due to non-vanishing scattering rate $\gamma$
are incorporated through the smearing of the critical region around
$\mu_c$. Note that the choice of the $\arctan$-function in the
$\mu$-dependence in Eq.~(\ref{gap-ansatz-main}) is suggested by
the $T=0$ gap equation (A11) in {\sl I}.
\end{itemize}

In total, we have three parameters, $\Delta_0$, $B_c$, and $\gamma$
for the description of the gap generation, which can be used to fit
the experimental data. In order to compare the theory and the
experiment, we substitute the gaps $\Delta_{\pm}$ given in
(\ref{gap-ansatz}) [with $\Delta(B,\mu_{\pm})$ from
Eq.~(\ref{gap-ansatz-main})] and dependences $\mu(V_g)$ and $E_Z(B)$
in Eqs.~(\ref{V2mu}) and (\ref{Zeeman-enh}) (extracted from the
experimental data) into the expression for the Hall conductivity:
\be \label{dcHall1} \sigma_{xy} =
\frac{1}{2}[\tilde{\sigma}_{xy}(\Delta_{+},\mu_+) +
\tilde{\sigma}_{xy}(\Delta_{-},\mu_-)]. \ee
The conductivity  $\tilde{\sigma}_{xy}$ in the limit $B \to \infty$
can be expressed in terms of the digamma function $\Psi(x)= d \ln
\Gamma(x)/dx$ (see Appendix~B in {\sl I}), namely
\begin{widetext}
\begin{equation}
\tilde{\sigma}_{xy}(\Delta_{\pm},\mu_{\pm})
=- \frac{2e^2{\rm sgn}(eB)}{\pi h}{\rm Im}\left\{
\Psi\left(\frac{\gamma_{\rm tr}+i(\mu_{\pm}+\Delta_{\pm})}{2\pi T}
+\frac{1}{2}\right)
-\frac{\gamma_{\rm tr}}{2\pi T}
\Psi^\prime\left(\frac{\gamma_{\rm tr} +i(\mu_{\pm}+\Delta_{\pm})}{2\pi
T}+\frac{1}{2}\right)+ (\Delta_{\pm} \to - \Delta_{\pm}) \right\},
\label{dcHall}
\end{equation}
\end{widetext}
where $\gamma_{\rm tr}$ is the transport scattering rate. One can
verify that for $T = \gamma_{\rm tr}= 0$ and $\Delta_{-} = \Delta_{+}
=\Delta$, Eq.~(\ref{dcHall1}) reduces to Eq.~(\ref{sigma_xy-ideal}).

\section{Results and their interpretation}
\label{sec:results}

\begin{figure}
\begin{center}
\includegraphics[width=.48\textwidth]{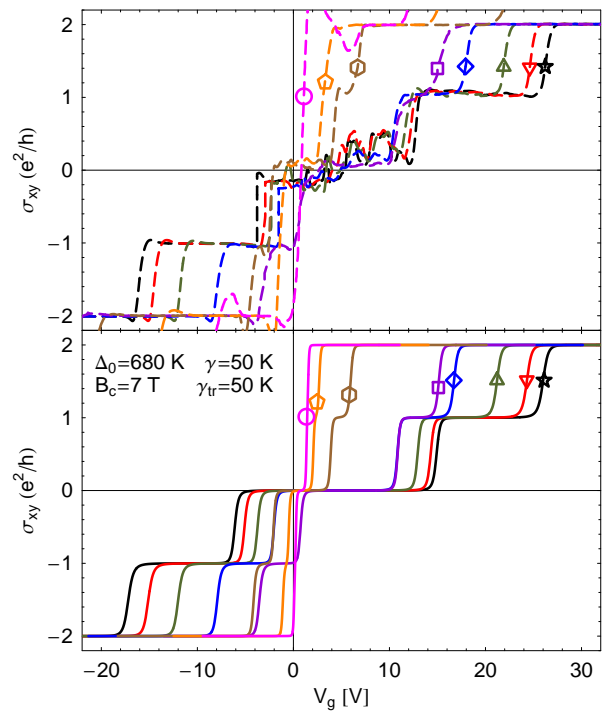}
\caption{(Color online) Hall conductivity from the data in Ref.~\onlinecite{Zhang2006PRL}
(upper panel) and in the theoretical model (lower panel) for
magnetic fields $B=9$~T (circle), $11.5$~T (pentagon), $17.5$~T
(hexagon), $25$~T (square), $30$~T (diamond), $37$~T (up triangle),
$42$~T (down triangle), and $45$~T (star). The parameters in the
model are $\gamma=\gamma_{\rm tr}=50$~K, and temperature $T
=30$ mK.} \label{sigmaHall-allB}
\end{center}
\end{figure}

The best fit is obtained while using the following values of the
parameters:
\begin{subequations}
\label{set}
\begin{align}
\Delta_0 &= 680~\mbox{K}, \label{Delta_0}\\
B_c &= 7~\mbox{T},\\
\gamma & = \gamma_{\rm tr}=50~\mbox{K}. \label{Gamma}
\end{align}
\end{subequations}
Note that these parameters capture essentially all the
non-perturbative physics of a strongly interacting model at hand.
Remarkably, it is sufficient to get a nearly perfect fit for a wide
range of the magnetic fields, see Fig.~\ref{sigmaHall-allB}. In the
upper panel of the Fig.~\ref{sigmaHall-allB}, we present the
experimental data for $\sigma_{xy}$, and in the lower panel their
description within our phenomenological model.

The theoretical curves plotted in Fig. \ref{sigmaHall-allB} at $T =
30 \, \mbox{mK}$ remain practically the same over a very wide range
of temperatures up to $T \lesssim   \Delta_0$ when $\gamma$ and
$\gamma_{\rm tr}$ do not depend on $T$. At present there are no
available experimental data for the temperature dependence of $\nu
=0,\pm 1$ and $\nu=\pm 4$ plateaus. Furthermore, to compare our
model with such experimental data at high temperatures one should
also take into account the dependence of $\gamma$ and $\gamma_{\rm
tr}$ on the temperature.

The large value of $\Delta_0$ in Eq.~(\ref{Delta_0}), which is
necessary to fit the data, and the simple estimate of the coupling
constant $g$ made below Eq.~(\ref{b}) clearly imply a
strong-coupling regime. The coupling constant is $g =e^2/\epsilon
\hbar v_F = \alpha c/\epsilon v_F$, where $\alpha \simeq 1/137$ is
the fine-structure constant. With air on one side of the graphene
plane and SiO$_2$ on the other, the unscreened dielectric constant
of the medium is estimated in Ref.~\onlinecite{Alicea2006PRB} to be
$\epsilon \approx 1.6 \epsilon_0$. This corresponds to $g \approx
1.37$. We see that once again this supports the arguments that a
strong-coupling regime considered in this paper is more plausible
than a weak-coupling one.

\begin{figure}
\begin{center}
\includegraphics[width=.45\textwidth]{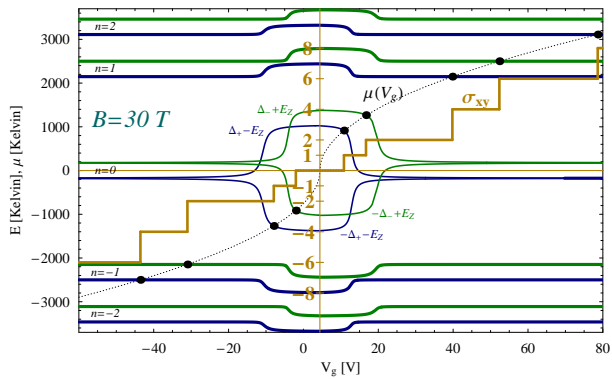}
\caption{(Color online) The theoretically reconstructed spectrum of one-particle
excitations in graphene at a fixed value of the magnetic field, $B=
30\, \mbox{T}$, as a function of the gate voltage. The energy levels
are shown by solid lines of different thickness which represents the
degeneracy of the Landau levels: thin and thick lines denote single
and double degeneracy, respectively. The dotted line shows the
dependence of the chemical potential on the gate voltage, see
Eq.~(\ref{V2mu}). The points of the intersection of the chemical
potential with the energy levels marks the position of the steps
between neighboring plateaus in the corresponding dependence of the
Hall conductivity on the voltage.} \label{spectrum}
\end{center}
\end{figure}

It is instructive to consider how the quasiparticle energy spectrum
is affected by the applied gate voltage in graphene. Such a spectrum
for a fixed value of the magnetic field is shown in Fig.~\ref{spectrum}.
In essence, it is the strong-coupling pairing
dynamics that is responsible for the drastic change of the energy
spectrum at low voltage. The four-fold degeneracy of the LL level is
lifted at low voltage because of the gap formation and the large Zeeman
splitting. When the voltage (and, therefore, the chemical potential and
the density) is large, the pairing plays no important role.

At large $|V_g|$, the LLL, which has the energy $E_0=0$ and the four-fold
degeneracy in absence of Zeeman splitting, splits into two levels with the
energies $E = \pm E_z$ and with the two-fold degeneracy (thick lines),
as shown in Fig.~\ref{spectrum}. When the absolute value of $\mu$ is
such that $\mu_- = \mu - E_Z$ is smaller than the critical value $\mu_c(B)$,
the gap $\Delta_{-}$ opens. This gap
causes the level with $E = E_Z$ to split into two nondegenerate levels
with energies $E = E_{Z} \pm \Delta_{-}$. Similar splitting of the energy
level $E= -E_Z$ into two levels with energies $E = -E_{Z} \pm \Delta_{+}$
is caused by a nonzero $\Delta_{+}$ when the value $\mu_+ = \mu + E_Z$
becomes smaller than $\mu_c(B)$. As should be clear from
Fig.~\ref{spectrum}, it is only the presence of such nondegenerate
levels in the energy spectrum of graphene that make the observation
of the $\nu = \pm 1$ plateaus possible. In approximately the whole
region where both gaps $\Delta_{\pm}$ are nonzero, the Hall conductivity
of graphene develops the $\nu=0$ plateau. Moreover, this correspondence
might be exact if the phase transition with respect to $V_g$ (or $\mu$)
is a strong first order phase transition.

It should be noted that the mechanism behind the creation of the $\nu =0,
\pm1$ plateaus in the strong-coupling regime is different from that at
weak coupling. In the latter case, the dominant term responsible for
the creation of the $\nu=0$ for small $\mu$ (or $V_g \sim V_0$) is the
excitonic gap $\Delta$, while the $\nu = \pm 1$ plateaus are related to
the Zeeman term (see the discussion after
Eq.~(\ref{E_n-gap}) above and Fig.~1 in
{\sl I}). In the present scenario with a strong-coupling dynamics,
there are two gaps, $\Delta_\pm$, and for $T =
\gamma_{\rm tr}= 0$ Eq.~(\ref{dcHall1}) reduces to
\begin{equation}
\label{sigma_xy-ideal-new}
\begin{split}
&\sigma_{xy}=-\frac{e^2}{h}{\rm sgn}(eB)\\
& \times\left[{\rm
sgn}(\mu_{+}) \theta(|\mu_{+}|-\Delta_{+})+{\rm
sgn}(\mu_{-})\theta(|\mu_{-}|-\Delta_{-})\right]
\end{split}
\end{equation}
instead of a more simple Eq.~(\ref{sigma_xy-ideal}).
For the case $E_Z < \Delta_{\pm}$ shown in Fig.~\ref{spectrum},
this dependence
of $\sigma_{xy}$ on $\mu$ implies that
while the size of the 0th plateau is
$\delta\mu_{0} =|\Delta_{+} + \Delta_{-} - 2E_Z|$, the size of
the $\pm 1$ plateaus is $\delta\mu_{1}=
|\Delta_{-} - \Delta_{+} + 2 E_Z|$.
When the effects of
non-vanishing widths $\gamma$ and $\gamma_{\rm tr}$ are taken into
account, the results become more complicated of course.
However, the main
qualitative features regarding the $\nu=0,\pm 1$ plateaus are
captured already by Eq.~(\ref{sigma_xy-ideal-new}). It is clear, for
example, that these plateaus result from a subtle interplay between
the gaps $\Delta_{\pm}$ and the Zeeman energy. We also note that the
$\nu=\pm1$ plateaus disappear when  $\Delta_{\pm}=0$. It is also
absent when $E_{Z}=0$ and $\Delta_{+} = \Delta_{-}$ (see also the
next section).

\section{Hall conductivity without enhanced Zeeman splitting}
\label{sec:noZeeman}

\begin{figure}
\begin{center}
\includegraphics[width=.48\textwidth]{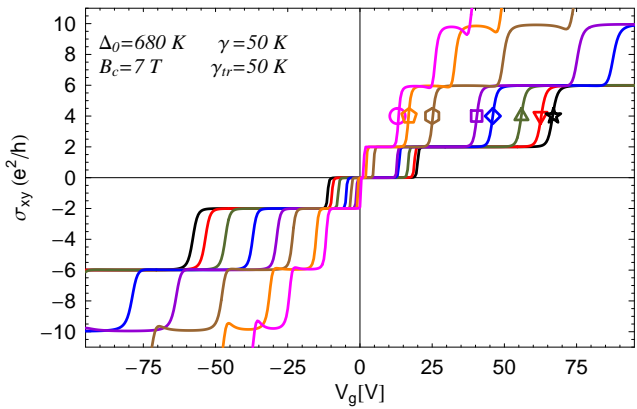}
\caption{(Color online) Hall conductivity in the case of a negligible Zeeman energy. As in
Fig.~\ref{sigmaHall-allB}, the symbols label different values of the magnetic
field: $B=9$~T (circle), $11.5$~T (pentagon), $17.5$~T (hexagon),
$25$~T (square), $30$~T (diamond), $37$~T (up triangle),
$42$~T (down triangle), and $45$~T (star).}
\label{fig:noZeeman}
\end{center}
\end{figure}

In order to appreciate the role of the enhanced Zeeman splitting, which
is required to fit the data in Ref.~\onlinecite{Zhang2006PRL},
it will be instructive
to consider also the case without Zeeman splitting, i.e., $E_{Z}=0$.
It should be
clear that this approximation should also describe well a system with an unenhanced
Zeeman energy $E_Z$, as given in Eq.~(\ref{Zeeman-usual}), when
$E_Z \alt \gamma$.
By making use of the same model parameters as in the previous section, see
Eq.~(\ref{set}), we plot the theoretical curve for the Hall conductivity as a
function of the gate voltage in Fig.~\ref{fig:noZeeman}.

To calculate the dependence of the Hall conductivity on the gate
voltage, here we used the expressions in Eqs.~(3.14) and (3.15) in
Ref.~\onlinecite{Gusynin2006PRB}. The corresponding result
includes the contributions of all Landau levels, which are needed in
order to reproduce not only the lowest plateaus with $|\nu|\leq 2$,
as in Fig.~\ref{sigmaHall-allB}, but also the higher ones.

As we see from Fig.~\ref{fig:noZeeman}, the plateaus with $\nu=\pm1$
and $\nu=\pm 4k$ (with integer $k$) do not appear when the Zeeman
energy is negligible (or $E_{Z}\alt \gamma$).
Of course, this conclusion
is hardly surprising in view of the correspondence between the energy
spectrum in graphene and the form of the Hall conductivity illustrated
in Fig.~\ref{spectrum}. When the Zeeman splitting of the Landau levels
disappears, the plateaus with $\nu=\pm1$, $\nu=\pm4$, $\nu=\pm8$, etc.
collapse into a point. In the language of symmetry, this picture
corresponds to a partial removing the degeneracy of the LLL, when
the $U(4)$ symmetry is broken down to the $U(2)_{a}\times U(2)_{b}$
(see Appendix C in {\sl I}).

\section{Conclusion}
\label{sec:concl}

In this paper, based on the experimental data of Ref.~\onlinecite{Zhang2006PRL},
we developed a model for the exciton condensation dynamics in graphene in
a strong magnetic field. The non-perturbative dynamics corresponds to a
strong coupling regime with a realistic value of the coupling constant
$g = e^2/\epsilon\hbar v_F$ of the Coulomb interaction.

On theoretical side, the model incorporates the main features of the
phenomenon of the magnetic catalysis\cite{Gusynin1995PRD} (the
generation of an excitonic gap in a strong magnetic field) and its
realization in graphene considered in
Refs.~\onlinecite{Khveshchenko2001PRL,Gorbar2002PRB}. On
phenomenological side, the model is based on features extracted from
the experimental data in Ref. \onlinecite{Zhang2006PRL}. Among them,
the relation between the gate voltage applied to the graphene device
and the chemical potential $\mu$ is particularly important. Another
important point established is a strong enhancement of the Zeeman
splitting (spin gap) in graphene in strong magnetic fields $B
\gtrsim 9 \,\mbox{T}$. The existing experimental data, however, do
not allow to extract the dependence of the spin gap on $T$ and $\mu$
and here we considered its dependence on the field $B$ only.

The enhanced Zeeman splitting alone does not allow to explain the
occurrence of the additional plateaus with the filling factors $\nu
=\pm1$ in magnetic fields $B\gtrsim 20 \,\mbox{T}$ observed in
Ref.~\onlinecite{Zhang2006PRL}. In addition to such a splitting, it
is necessary to remove the sublattice degeneracy in graphene to
explain the origin of these plateaus. The available experimental
data\cite{Zhang2006PRL} already contain a lot of constraints on
possible microscopical mechanism which removes this degeneracy. In
addition to this, the observation of the $\nu=0$ plateau at rather
high temperatures\cite{Geim-private} indicates that a strong
coupling regime is more believable than the weak coupling regime.

Still further experiments are necessary  to establish the
temperature evolution of the $\nu =0, \pm1, \pm 4$
plateaus and their sensitivity to the quality of the samples. This
will hopefully allow to decide which of the
mechanisms mentioned in the Introduction is realized
in graphene in a strong magnetic field. For example, there is a
suggestion how to detect a gap either in
microwave\cite{Gusynin2006PRL} or optical\cite{Gusynin2006}
response. Recent measurements done in the far infrared
region\cite{Sadowski2006,Li2006PRB} show that the second method may
work if the measurements will be done not in epitaxial graphite as
in Ref.~\onlinecite{Sadowski2006} or in highly oriented pyrolytic
graphite as in Ref.~\onlinecite{Li2006PRB} but in graphene in a
strong magnetic field.

\acknowledgments

We are grateful to P. Kim and Y. Zhang for sending us original
experimental data of their work\cite{Zhang2006PRL} and to A.K. Geim
for discussing with us the unpublished results. Useful discussions
with J.P. Carbotte are acknowledged. The work of V.P.G. was
supported by the SCOPES-project IB7320-110848 of the Swiss NSF
and by Ukrainian State Foundation for Fundamental Research. The
work of V.A.M. was supported by the Natural Sciences and Engineering
Research Council of Canada. He acknowledges the warm hospitality
of Prof. Taichiro Kugo and Prof. Teiji Kunihiro during his stay
at Yukawa Institute for Theoretical Physics.
The work of S.G.S. was supported by the
Natural Sciences and Engineering Research Council of Canada and the
Canadian Institute for Advanced Research.


\begin{thebibliography}{99}
\bibitem{foot1} In the both cases, the $SU(4)$ is associated with
spin-pseudospin degrees of freedom, but while in graphene pseudospin
is related to the sublattice index, in bilayer quantum Hall systems
pseudospin is related to the layer index.
\end{thebibliography}
\end{document}